\documentclass[journal, draftcls, onecolumn, twoside, 12pt]{IEEEtran}
\usepackage{amssymb, amsmath, amsfonts}
\usepackage[mathscr]{eucal}
\usepackage{url, hyperref}
\newcommand{\tr}{\textrm{tr}}
\newcommand{\zz}{{\mathbb Z}}
\newcommand{\x}{{\mathbf{x}}}

\newcommand{\cc}{{\mathbb C}}
\newtheorem{lem}{Lemma}[section]
\newtheorem{defn}{Definition}[section]
\newtheorem{theorem}{Theorem}[section]
\newtheorem{propos}{Proposition}[section]
\newtheorem{corollary}[theorem]{Corollary}
\newtheorem{remark}{Remark}

\title{Invariance of Stationary and Ergodic Properties of a Quantum Source under Memoryless Transformations}%\thanks{This work was
\author{Alexei Kaltchenko$^{1}$\footnote{e-mail: \href{mailto:akaltche@ece.uwaterloo.ca}{akaltche@ece.uwaterloo.ca}} \; and \; En-Hui Yang$^{2}$ \\  $^{1,2}$ {\footnotesize E\&CE Department, University of Waterloo,} \\
{\footnotesize Waterloo, Ontario N2L 3G1, Canada} \\}

\begin{document}
\maketitle
\begin{abstract}
We prove that the stationarity and the ergodicity of a quantum
source $\{\rho_m \}_{m=1}^{\infty}$ are preserved by any
trace-preserving completely positive linear map of the tensor
product form ${\cal E}^{\otimes m}$, where a copy of ${\cal E}$
acts locally on each spin lattice site. We also establish
ergodicity criteria for so called classically-correlated quantum
sources.
\end{abstract}
\section{Introduction}
The quantum ergodicity proves to be as instrumental in studying
quantum information systems as is the classical ergodicity in
studying classical systems. To give a rough idea of the role that
quantum ergodicity plays in quantum information theory, one may
name just one result, the quantum extension\cite{Q-McMillan} of
Shannon-McMillan theorem.

In this paper we are concerned with stationary and ergodic
properties of quantum sources. Specifically, we study the case
when a stationary and ergodic (weakly mixing or strongly mixing,
respectively) quantum source $\{\rho_m \}_{m=1}^{\infty}$ is
subjected to a trace-preserving completely positive linear
transformation (map) of the tensor product form ${\cal E}^{\otimes
m}$, where a copy of ${\cal E}$ locally acts on each spin lattice
site. We present several technical lemmas and prove that the map
preserves all the listed source properties. Such maps describe the
effect of a transmission via a {\em memoryless channel} as well as
the effect of {\em memoryless coding}, both lossless and lossy
ones. As a corollary of our main result, we also establish
ergodicity criteria for so called {\em classically-correlated}
quantum sources.
\section{Quantum Sources: Mathematical Formalism and
Notation}\label{Notation} Before we define a general quantum
source, we give an informal, intuitive definition of a so-called
{\em classically correlated} quantum source  as  a
triple\cite{king} consisting of {\em quantum messages}, a {\em
classical probability distribution} for the messages, and the {\em
time shift}. Such the triple uniquely determines a state of a
one-dimensional quantum lattice system. If quantum-mechanical
correlation between the messages exists, one gets the notion of a
general quantum source. While any given state corresponds to
infinitely many different quantum sources, the quantum state
formalism essentially captures all the information-theoretic
properties of a corresponding quantum source. Thus, the notion of
"quantum source" is usually identified with the notion of "state"
of the corresponding lattice system and used interchangeably.

Let~$Q$ be an infinite quantum spin lattice system over lattice
$\mathbb{Z}$ of integers.  To describe $Q$, we use the standard
mathematical formalism introduced in~\cite[Sec.~6.2.1]{bratteli-2}
and~\cite[Sec.~1.33 and Sec.~7.1.3]{ruelle} and borrow notation
from~\cite{Q-McMillan} and~\cite{mosonyi}. Let ${\mathfrak A}$ be
a $C^{*}$-algebra\footnote{The algebra of all bounded linear
operators may be simply thought of as the algebra of all square
matrices with the standard matrix operations including
conjugate-transpose.} with identity of all bounded linear
operators ${\cal B}({\cal H})$ on a $d$-dimensional Hilbert space
${\cal H}$, $d < \infty$. To each $\x \in \zz$ there is associated
an algebra ${\mathfrak A}_{\mathbf{x}}$ of observables for a spin
located at site $\x$, where  ${\mathfrak A}_{\mathbf{x}}$ is
isomorphic to ${\mathfrak A}$ for every $\x$. The local
observables in any finite subset $\Lambda\subset\zz$ are those of
the finite quantum system
$$
{\mathfrak A}_{\Lambda}:= \bigotimes_{\x \in \Lambda} {\mathfrak
  A}_{\x}
$$
The quasilocal algebra ${\mathfrak A}_{\infty}$ is the operator
norm completion of the normed algebra $\bigcup\limits_{\Lambda
\subset \zz}{\mathfrak A}_\Lambda $, the union of all local
algebras ${\mathfrak A}_\Lambda$ associated with finite
$\Lambda\subset\zz$. A state of the infinite spin system is given
by a normed positive functional $$\varphi \ : \ {\mathfrak
A}_{\infty} \to \cc.$$ We define a family of states
$\{\varphi^{(\Lambda)}\}_{\Lambda \subset \zz}$, where
$\varphi^{(\Lambda)}$ denotes the restriction of the state
$\varphi$ to a finite-dimensional subalgebra ${\mathfrak
A}_\Lambda$, and assume that $\{\varphi^{(\Lambda)}\}_{\Lambda
\subset \zz}$ satisfies the so called {\em consistency}
condition{\cite{Q-McMillan, king}, that is
\begin{equation}\label{consistency}
 \varphi^{ (\Lambda)}= \varphi^{(\Lambda^{'})}
\upharpoonright{\mathfrak A}_{\Lambda} \end{equation}
 for any $\Lambda \subset
\Lambda^{'}.$ The consistent family
$\{\varphi^{(\Lambda)}\}_{\Lambda \subset \zz}$ can be thought of
as a quantum-mechanical counterpart of a consistent family of
cylinder measures. Since there is one-to-one correspondence
between the state $\varphi$ and the family
$\{\varphi^{(\Lambda)}\}_{\Lambda \subset \zz}$, any physically
realizable transformation of the infinite system~$Q$, including
coding and transmission of quantum messages, can be well
formulated using the states $\varphi^{ (\Lambda)}$ of finite
subsystems. Where the subset $\Lambda \in \mathbb{Z}$ needs to be
explicitly specified, we will use the notation $\Lambda(n)$,
defined as
\begin{equation*}
\Lambda(n):= \bigl\{ x \in \mathbb{Z}: x \in \{1,\ldots,n\}
\bigr\}
\end{equation*}

Let $\gamma$ (or $\gamma^{-1}$, respectively) denote a
transformation on ${\mathfrak A}_{\infty}$ which is induced by the
right (or left, respectively) shift   on the set $\zz$. Then, for
any $l \in \mathbb{N}$, $\gamma^{l}$ (or $\gamma^{-l}$,
respectively) denotes a composition of $l$ right (or left,
respectively) shifts. Now we are equipped to define the notions of
stationarity and ergodicity of a quantum source.
\begin{defn}\label{stationarity}
A state $\varphi$ is called $N$-{\it stationary} for an integer
$N$ if $\varphi \circ {\gamma}^N=\varphi$. For $N=1$, an $N$-{\it
stationary} state is called {\it stationary}.
\end{defn}
\begin{defn}
A $N$-stationary state is called $N$-{\it ergodic} if it is an
extremal point in the set of all $N$-stationary states. For $N=1$,
$N$-ergodic state is called ergodic.
\end{defn}
The following lemma which provides a practical method of
demonstrating the ergodicity of a state is due
to~\cite[Propos.~6.3.5,~Lem.~6.5.1]{ruelle}.
\begin{lem}
The following conditions are equivalent:
\begin{enumerate}
\item[(a)] A stationary state $\varphi$ on ${\mathfrak A}_{\infty}$ is
ergodic.
\item[(b)] For all~$a, b \in {\mathfrak A}_{\infty}$, it holds
\begin{equation}\label{ergodic} \lim_{n \to \infty}
\frac{1}{n} \sum_{i=1}^n \varphi(a \ {\gamma}^i(b)) = \varphi(a) \
\varphi(b).
\end{equation}
\item[(c)] For every selfajoint~$a \in {\mathfrak A}_{\infty}$, it
holds
\begin{equation*}
\lim_{n \to \infty} \varphi\Biggl( \biggl(\frac{1}{n} \sum_{i=1}^n
{\gamma}^i(a)\biggr)^2 \Biggr) = \varphi^2(a).
\end{equation*}
\end{enumerate}
%where ${\gamma}^i$ denotes the right $i$-shift $\underbrace
%{\gamma \circ \cdots \circ \gamma}_i$.
\end{lem}
Now we state a series of definitions\cite{bratteli-1} which
provide "stronger" notions of ergodicity:
\begin{defn}
A state is called {\it completely ergodic} if it is $N$-ergodic
for every integer $N$.
\end{defn}
\begin{defn}
A stationary state $\varphi$ on ${\mathfrak A}_{\infty}$ is called
weakly mixing if
\begin{equation}\label{weakly-mixing}
\lim_{n \to \infty} \frac{1}{n} \sum_{i=1}^n \left| { \varphi(a \
{\gamma}^i(b)) -\varphi(a) \ \varphi(b) } \right| =0, \quad
\forall a, b \in {\mathfrak A}_{\infty}.
\end{equation}
\end{defn}

\begin{defn}
A stationary state $\varphi$ on ${\mathfrak A}_{\infty}$ is called
strongly mixing if
\begin{equation}\label{stronly-mixing}
\lim_{i \to \infty}   \varphi(a \ {\gamma}^i(b)) =\varphi(a) \
\varphi(b), \quad \forall a, b \in {\mathfrak A}_{\infty}.
\end{equation}
\end{defn}
It is straightforward to see that~\eqref{stronly-mixing}
$\Rightarrow$~\eqref{weakly-mixing}~$\Rightarrow$~\eqref{ergodic}.
%%%%%%%%%%%%%%%%%%%%%%%%%%%%%%%%%%%%%%%%%%%%%%%%

Let $\textrm{tr}_{{\mathfrak A}_{\Lambda}}(\cdot)$ denote the
canonical trace on ${\mathfrak A}_{\Lambda}$ such that
$\textrm{tr}_{{\mathfrak A}_{\Lambda}}(e) =1$ for all
one-dimensional projections~$e$ in~${\mathfrak A}_{\Lambda}$.
Where an algebra on which the trace is defined is clear from the
context, we will omit the trace's subscript and simply write
$\textrm{tr}(\cdot)$. For each $\varphi^{(\Lambda)}$ there exists
a unique density operator $\rho_{\Lambda}\in {\mathfrak
A}_{\Lambda}$, such that
$\varphi^{(\Lambda)}(a)=\textrm{tr}(\rho_{\Lambda}a),\
a\in{\mathfrak A}_{\Lambda}$. Thus, any stationary state~$\varphi$
is uniquely defined by the family of density operators
$\{\rho_{\Lambda(m)} \}_{m=1}^{\infty}$. Where no confusion
arises,  we will use the following abbreviated notation for the
rest of the paper. For all $n \in \mathbb{N}$,
\begin{alignat*}{2}
& {\mathfrak A}^{(n)} &\ := &\ {\mathfrak A}_{\Lambda(n)}\\
& \psi^{(n)} & := &\ \psi^{\Lambda(n)}\\
& \rho_n &:=  &\ \rho_{\Lambda(n)}
\end{alignat*}

%%%%%%%%%%%%%%%%%%%%%%%%%%%%%%%%%%%%%%%%%%%%%%%%%%%%%%%%%%%%%
%%%%%%%%%%%%%%%%%%%%%%%%%%%%%%%%%%%%%%%%%%%%%%%%%%%%%%%%%%%%%
\section{Main Result}\label{Ergodicity-Invariance}%
In this section we present a sequence of lemmas and a theorem
which help to establish the ergodicity of a state. But first we
shall reformulate the stationary ergodic properties of an infinite
spin lattice system in terms of its finite subsystems. By
rewriting the consistency condition~\eqref{consistency},
Definition~\ref{stationarity}, and the
equations~(\ref{ergodic}--\ref{stronly-mixing}) in terms of
density operators, we obtain the following three elementary
lemmas\footnote{In what follows we abusively use the same symbol
to denote both an operator (or superoperator), confined to a
lattice box $\Lambda(m)$, and its "shifted" copy, confined to a
box $\{1+j,\ldots,\ m+j\}$, where the value of integer $j$ will be
understood from the context.}.
\begin{lem}\label{consistency-operator-form}
A family $\{\rho_{m} \}_{m=1}^{\infty}$ on ${\mathfrak
A}_{\infty}$ is consistent if and only if, for all positive
integers~$m, i<\infty$ and every $a \in {\mathfrak A}^{(m)}$, the
following holds:
\begin{equation}\label{consistency-operator-form-equation}
\tr(\rho_{m} \ a) = \tr\bigl(\rho_{m+i} \ (a \otimes I^{\otimes
i})\bigr),
\end{equation}
where $I^{\otimes i}$ stands for the $i$-fold tensor product of
the identity operators acting on respective spins.
\end{lem}
\begin{lem}\label{stationarity-operator-form}
A quantum source $\{\rho_{m} \}_{m=1}^{\infty}$ on ${\mathfrak
A}_{\infty}$ is stationary if and only if, for all positive
integers~$m, i<\infty$ and every $a \in {\mathfrak A}^{(m)}$, the
following equality is satisfied:
\begin{equation}
\tr(\rho_{m} \ a) = \tr\bigl(\rho_{m+i} \ (I^{\otimes i} \otimes
a)\bigr),
\end{equation}
\end{lem}
\begin{lem}\label{finite-ergodicity}
A stationary quantum source $\{\rho_{m} \}_{m=1}^{\infty}$ on
${\mathfrak A}_{\infty}$ is ergodic (weakly mixing or strongly
mixing, respectively) if and only if, for every positive
integer~$m <\infty$ and all $a, b \in {\mathfrak A}^{(m)}$, the
equality \eqref{ergodic1} $\bigl($\eqref{weakly-mixing1} or
\eqref{stronly-mixing1}, respectively$\bigr)$ holds:
\begin{align}
&\lim_{n \to \infty} \frac{1}{n} \sum_{i=m}^n \tr\bigl(\rho_{m+i}
\ (a \otimes I^{\otimes (i-m)} \otimes b)\bigr) = \tr(\rho_{m} a)
\ \tr(\rho_{m}
b), \label{ergodic1}  \\
&\lim_{n \to \infty} \frac{1}{n} \sum_{i=m}^n
\bigl|\tr\bigl(\rho_{m+i} \ (a \otimes I^{\otimes (i-m)} \otimes
b)\bigr) - \tr(\rho_{m} a) \ \tr(\rho_{m} b)\bigr|
=0, \label{weakly-mixing1}\\
&\lim_{i \to \infty} \tr\bigl(\rho_{m+i} \ (a \otimes I^{\otimes
(i-m)} \otimes b)\bigr) = \tr(\rho_{m}  a) \ \tr(\rho_{m}
b),\label{stronly-mixing1}
\end{align}
\end{lem}
We now need to fix some additional notation. Let ${\cal E}$ be an
arbitrary trace-preserving quantum operation that has the input
space ${\cal B}({\cal H})$. Without loss of generality we assume
that the output space for ${\cal E}$ is also ${\cal B}({\cal H})$.
It is known\cite{kraus} that ${\cal E}$ is a trace-preserving
completely positive linear (TPCPL) map. Next, we define a
composite map%
\[{\cal E}^{\otimes m} \ : \ {\mathfrak
A}^{(m)} \to {\mathfrak A}^{(m)}, \qquad \forall m >
0.\]%
We point out that such a tensor product map is the most
general description  of a {\em quantum memoryless
channel}\cite{barnum}.
\begin{theorem}\label{main}
If $\{\rho_{m} \}_{m=1}^{\infty}$ is a stationary and ergodic
(weakly mixing or strongly mixing, respectively) source, then so
is  $\left\{ { {\cal E}^{\otimes m}\bigl(\rho_{m}\bigr) }
\right\}_{m=1}^{\infty}$. The proof of this theorem is given in
the appendix~\ref{Proof-of-erg-theorem}.
\end{theorem}
\begin{remark}
We note that any weakly or strongly mixing quantum source is also
completely ergodic. Then, for such sources, the theorem trivially
extends to cover TPCPL maps of the form ${\bigl({\cal
E}^k\bigr)}^{ \otimes (m/k)}$, $(m/k) \in \zz$, where ${\cal E}^k$
acts on $k$-blocks of lattice, in direct analogy with a $k$-block
classical coding. Thus, our work is the quantum generalization of
a well-known classical information-theoretic
result\cite[chap.~7]{berger} for memoryless- and block-coding and
channel transmission.
\end{remark}
\begin{defn}\label{classically-correlated}
We define a classically correlated quantum source
$\{\rho_{m}^{cls} \}_{m=1}^{\infty}$ by an equation
\begin{equation}
\rho_{m}^{cls}: = \sum_{x_1 ,x_2 , \ldots ,x_m} p(x_1 ,x_2 ,
\ldots ,x_m )|x_1 \rangle \langle x_1 | \otimes |x_2 \rangle
\langle x_2 | \otimes \cdots  \otimes |x_m \rangle \langle x_m |,
\end{equation}
where $p(\cdot)$ stands for a probability distribution, and for
every $i$, $|x_i \rangle$ belongs to some fixed
linearly-independent set $S: = \left\{ {|\psi _1 \rangle ,|\psi _2
\rangle , \ldots ,|\psi _d \rangle } \right\}$ of vectors in the
Hilbert space ${\cal H}$. We recall that ${\cal H}$ is the support
space for the operators in ${\mathfrak A}$. The set $S$ is
sometimes called a {\em quantum alphabet}.
\end{defn}
\begin{corollary}\label{crllry}
If a classical probability distribution~$p(\cdot)$ in
Definition~\ref{classically-correlated} is a stationary and
ergodic (weakly mixing or strongly mixing, respectively), then so
is the quantum source $\{\rho_{m}^{cls} \}_{m=1}^{\infty}$. The
proof of this corollary is given in the
appendix~\ref{Proof-of-Corollary}.
\end{corollary}

\appendices
\section{Conditional expectation}
Let $\tilde {\mathfrak A}$ be a $C^*$-subalgebra of~${\mathfrak
A}$, and let $E : {\mathfrak A} \rightarrow \tilde{\mathfrak A}$
be a linear mapping which sends the density of every state
$\varphi$ on ${\mathfrak A}$ to the density of the state~$\varphi
\upharpoonright \tilde{\mathfrak A}$. Such a mapping is usually
called a {\em conditional expectation} and has the following
properties\cite[Propos.~1.12]{ohya}:
\begin{enumerate}
\item[(a)] if $a \in {\mathfrak A}$ is positive operator,
then so is $E (a) \in \tilde{\mathfrak A}$;
\item[(b)] $E (b) = b$ for every $b \in \tilde{\mathfrak A}$;
\item[(c)] $E (ab) = E (a) b$ for every~$a \in {\mathfrak
A}$ and~$b\in \tilde{\mathfrak A}$;
\item[(d)] for every~$a \in {\mathfrak A}$, it holds \[\tr_{{\mathfrak A}}(a) = \frac{\tr_{{\mathfrak A}}(I)}{\tr_{\tilde{\mathfrak
A}}(I)} \tr_{\tilde{\mathfrak A}}\bigl(E (a)\bigr),  \] where $I$
stands for identity operator.
\end{enumerate}
%%%%%%%%%
%%%%%%%%%%

%%%%%%%%%%%%%%%%%%%%%%%%%%%%%%%%%%%%%%%%%%%%%%%%%%%%%%%%%%%%%%%
%%%%%%%%%%%%%%%%%%%%%%%%%%%%%%%%%%%%%%%%%%%%%%%%%%%%%%%%%%%%%
\section{Proof of Theorem \ref{main}}\label{Proof-of-erg-theorem}%
%\begin{proof}
%\textbf{Proof of Theorem \ref{main}:}\\
For any TPCPL map there exists a so-called "operator-sum
representation"\cite{barnum},\cite{kraus}. Then, an $m$-fold
tensor product map ${\cal E}^{\otimes m}$ has the following
representation:
\begin{equation}\label{operation-representation1}
{\cal E}^{\otimes m}\bigl(\rho_{m}\bigr)
=\sum_{j_1,j_2,\ldots,j_m} \bigl(A_{j_1} \otimes A_{j_2}
\otimes\cdots\otimes A_{j_m} \bigr) \rho_{[1,m]} \bigl(A_{j_1}
\otimes A_{j_2} \otimes\cdots\otimes A_{j_m}
\bigr)^{\dag}\end{equation}%
with
\begin{equation}\label{operation-representation2}
\sum_{i} A_i^{\dag} A_i = I, \quad A_{i}, I \in {\mathfrak A},
\end{equation}%
where $I$ stands for identity operator.\\
Due to~\eqref{operation-representation1} and
\eqref{operation-representation2}, the following three equalities
hold for all positive integers~$m <i <\infty$ and all $a, b \in
{\mathfrak A}^{(m)}$
\begin{equation}\label{set}
\begin{gathered}
\tr\bigl({\cal E}^{\otimes (m+i)}(\rho_{m+i}) \ (a \otimes
I^{\otimes (i-m)} \otimes b)\bigr) = \tr\bigl( \rho_{m+i} \ (
\tilde a \otimes I^{\otimes (i-m)} \otimes
\tilde b )\bigr),\hfill \\
\tr({\cal E}^{\otimes m}(\rho_{m})  a) =
\tr(\rho_{m}  \tilde a),\hfill \\
\tr({\cal E}^{\otimes m}(\rho_{\Lambda(m)})  b) =
\tr(\rho_{m}  \tilde b),\hfill \\
\end{gathered}
\end{equation} where $a, b \in {\mathfrak A}^{(m)}$ and $\tilde a$
and $\tilde b$ are defined as follows:
\begin{gather*}
\tilde a := \sum_{j_1,j_2,\ldots,j_m} \bigl(A_{j_1} \otimes
A_{j_2} \otimes\cdots\otimes A_{j_m} \bigr)^{\dag} a \bigl(A_{j_1}
\otimes A_{j_2} \otimes\cdots\otimes A_{j_m}
\bigr), \\
\tilde b :=  \sum_{j_1,j_2,\ldots,j_m} \bigl(A_{j_1} \otimes
A_{j_2} \otimes\cdots\otimes A_{j_m} \bigr)^{\dag} b \bigl(A_{j_1}
\otimes A_{j_2} \otimes\cdots\otimes A_{j_m}
\bigr). \\
\end{gather*}
Combining~\eqref{set} with Lemma~\ref{finite-ergodicity}, we
obtain the ergodicity (weakly mixing or strongly mixing,
respectively) of $\left\{ { {\cal E}^{\otimes
m}\bigl(\rho_{m}\bigr) } \right\}_{m=1}^{\infty}$. In a similar
manner, the application of Lemma~\ref{consistency-operator-form}
establishes  the consistency of $\left\{ { {\cal E}^{\otimes
m}\bigl(\rho_{m}\bigr) } \right\}_{m=1}^{\infty}$, and the
application of Lemma~\ref{stationarity-operator-form} establishes
the stationarity of $\left\{ { {\cal E}^{\otimes
m}\bigl(\rho_{m}\bigr) } \right\}_{m=1}^{\infty}$.
%\end{proof}
$\blacksquare$
%%%%
%
\section{Proof of Corollary \ref{crllry}}\label{Proof-of-Corollary}
%\begin{proof}%
Let $S_{\bot}:= \left\{ {|e _1 \rangle ,|e _2 \rangle , \ldots ,|e
_d \rangle } \right\}$ be any orthonormal basis in~${\cal H}$, and
let $\{\tilde\rho_{m}^{cls} \}_{m=1}^{\infty}$ be the source with
alphabet $S_{\bot}$ and distribution $p(\cdot)$. For
$i=1,\ldots,d$, we define a set $\{A_i\}$ of linear operators as
follows
\begin{equation}
A_i:=|\psi _i \rangle \langle e_i|.
\end{equation}
Then, set $\{A_i\}$ satisfies \eqref{operation-representation2},
and we define a TPCPL map ${\cal E}^{\otimes m}$ as
in~\eqref{operation-representation1}. Consequently, we have
$\bigl(\rho_{m}^{cls}\bigr) = {\cal E}^{\otimes m}\bigl(\tilde
\rho_{m}^{cls}\bigr)$. Thus, to complete the proof, we need to
show that $\{\tilde \rho_{m}^{cls} \}_{m=1}^{\infty}$ on
${\mathfrak A}_{\infty}$ is ergodic (weakly mixing or strongly
mixing, respectively). Let ${\mathfrak C}$ be a subalgebra of
${\mathfrak A}$ spanned by the set $\{ |e_i\rangle \langle e_i| \
: \ |e_i\rangle \in S_{\bot} \}$. We extend ${\mathfrak C}$ to a
quasilocal algebra ${\mathfrak C}_{\infty} \subset {\mathfrak
A}_{\infty}$ over lattice $\zz$ in the same way we did for
${\mathfrak A}_{\infty}$. The algebra ${\mathfrak C}_{\infty}$ is
abelian due to the orthogonality of the set $S_{\bot}$.
%%%%%%%%%%%%%%%%%%%%%%%%%%%%%%%%%%%%%%%%%%%%%%%
%%%%%%%%%%%%%%%%%%%%%%%%%%%%%%%%%%%%%%%%%%%%%%%%
For any integer $m>1$, let $E_m : {\mathfrak A}^{(m)} \rightarrow
{\mathfrak C}^{(m)}$ denote the conditional expectation.
Since~${\mathfrak C}^{(m)}$ is a maximal abelian subalgebra
of~${\mathfrak A}^{(m)}$, we have $\tr_{{\mathfrak A}^{(m)}}(I) =
\tr_{{\mathfrak C}^{(m}}(I)$. Moreover, by our construction,
$\tilde \rho_{m}^{cls}$ is an element of algebra ${\mathfrak
C}^{(m)} \subset {\mathfrak A}^{(m)}$ for every $m$. Then, the
following equalities hold by the properties of conditional
expectation for all positive integers $m < i < \infty$ and all $a,
b \in {\mathfrak A}^{(m)}$:
\begin{align*}
&\tr_{{\mathfrak A}^{(m+i)}} \Bigl(\tilde \rho^{cls}_{m+i} \bigl(a\otimes I^{\otimes(i-m)}\otimes b\bigr) \Bigr)%
= \tr_{{\mathfrak C}^{(m+i)}}\bigg( E_{m+i}\Bigl(  \tilde \rho^{cls}_{m+i} \bigl(a\otimes I^{\otimes(i-m)}\otimes b\bigr) \Bigr) \bigg)\\%
= \  &\tr_{{\mathfrak C}^{(m+i)}}\Big( \tilde \rho^{cls}_{m+i} E_{m+i}\bigl(a\otimes I^{\otimes(i-m)}\otimes b\bigr) \Big),  \\
&\tr_{{\mathfrak A}^{(m)}} (\tilde \rho^{cls}_m a) =
\tr_{{\mathfrak C}^{(m)}} \bigl(E_m(\tilde \rho^{cls}_m a)\bigr) =
\tr_{{\mathfrak C}^{(m)}}
\bigl(\tilde \rho^{cls}_m E_m(a)\bigr),\\
&\tr_{{\mathfrak A}^{(m)}} (\tilde \rho^{cls}_m b) =
\tr_{{\mathfrak C}^{(m)}} \bigl(E_m(\tilde \rho^{cls}_m b)\bigr) =
\tr_{{\mathfrak C}^{(m)}} \bigl(\tilde \rho^{cls}_m E_m(b)\bigr).
\end{align*}
Thus, if $\{\tilde \rho_{m}^{cls} \}_{m=1}^{\infty}$ is
consistent, stationary, and ergodic (weakly mixing or strongly
mixing, respectively) on ${\mathfrak C}_{\infty}$, then it also
holds on ${\mathfrak A}_{\infty}$ by the
lemmas~\ref{consistency-operator-form},
\ref{stationarity-operator-form}, and~\ref{finite-ergodicity}.
%%%%%%%%%%%%%%%%%%%%%%%%%%%%%%%%%%%%%%%%%%%%%%%%
%%%%%%%%%%%%%%%%%%%%%%%%%%%%%%%%%%%%%%%%%%%%%%%%%%
Finally, we note that since ${\mathfrak C}_{\infty}$ is abelian,
$\{\tilde \rho_{m}^{cls} \}_{m=1}^{\infty}$ on ${\mathfrak
C}_{\infty}$ is ergodic (weakly mixing or strongly mixing,
respectively) if and only if so is~$p(\cdot)$ by
Proposition~\ref{abelian->classic} from the appendix.
$\blacksquare$
%\end{proof}
%
%%%
\section{States on Quasilocal Commutative
$C^*$-algebras}\label{commutative-algebras}%
Let $\mathfrak{B}$ be an arbitrary commutative $k$-dimensional
$C^*$-subalgebra of~${\cal B}({\cal H})$, and let
$\mathfrak{B}_{\infty}$ be a quasilocal algebra
$\mathfrak{B}_{\infty}$ over lattice $\mathbb{Z}$ with local
algebras $\mathfrak{B}_{\x}$ isomorphic to $\mathfrak{B}$ for
every $\x \in \mathbb{Z}$, i.e., $\mathfrak{B}_{\infty}$ is
constructed in the same way as is ${\mathfrak A}_{\infty}$ in
Section~\ref{Notation}. Then, for any $\Lambda \subset
\mathbb{Z}$, every minimal projector in $\mathfrak{B}_{\Lambda}$
is necessarily one-dimensional, and the density operator for every
pure state $\varphi^{(\Lambda)}$ on $\mathfrak{B}_{\Lambda}$ is
exactly a one-dimensional projector. Let $\bigl\{ |z_i\rangle
\langle z_i| \bigr\}_{i=1}^{k}$ be a collection of the density
operators for all the distinct pure states on $\mathfrak{B}$. We
then define an abstract set $\mathcal{Z}:=\{ z_i \}_{i=1}^{k}$,
where every element $z_i$ symbolically corresponds to the
operator~$|z_i\rangle \langle z_i| $, and $z_i \neq z_j$ for all
$i \neq j$. For every finite lattice subset $\Lambda \in
\mathbb{Z}$, we define the Cartesian product
$$
{\mathcal{Z}}^{\Lambda}:=  \underset{\x \in
\Lambda}{\pmb{\times}}{\mathcal{Z}}_{\x},
$$
i.e., the elements $\omega$ of ${\mathcal{Z}}^{\Lambda (n)}$ have
the form $\omega = \omega_1\ldots\omega_{n}$, $\omega_i \in
\mathcal{Z}$. It is easy to see that, for every $\Lambda \in
\mathbb{Z}$, the set ${\mathcal{Z}}^{\Lambda}$ and the set of all
one-dimensional projectors in $\mathfrak{B}_{\Lambda}$ are in
one-to-one correspondence: $\omega \longleftrightarrow
|\omega\rangle \langle \omega|$. Consequently, there is one-to-one
correspondence between the set of all projectors in
$\mathfrak{B}_{\Lambda}$ and ${\mathscr{P}}^{\Lambda}(
{\mathcal{Z}})$, the Cartesian product of the power sets of
${\mathcal{Z}}$. In particular,  every projector $p \in
\mathfrak{B}_{\Lambda}$ corresponds to a set $\bigl\{ \omega :
\omega \in {\mathcal{Z}}^{\Lambda}, |\omega\rangle \langle \omega|
\leqslant p \bigr\}$. We note that, equipped with the product of
the discrete topologies of the sets ${\mathcal{Z}}_{\x}$,
${\mathcal{Z}}^{\Lambda}$ is a compact space, and the pair
$\bigl({\mathcal{Z}}^{\Lambda}, {\mathscr{P}}^{\Lambda}(
{\mathcal{Z}}) \bigr)$ defines a measurable space. Thus, by
Gelfand-Naimark theorem\cite[Chap.~11]{rudin2} and Riesz
representation theorem\cite[Sec.~2.14]{rudin1}, for any pure or
mixed state $\varphi^{(\Lambda)}$ on $\mathfrak{B}_{\Lambda}$,
there exists a unique positive measure on
$\bigl({\mathcal{Z}}^{\Lambda}, {\mathscr{P}}^{\Lambda}(
{\mathcal{Z}}) \bigr)$, denoted by $\mu_{\Lambda}$, such that the
following equality holds for any projector $p \in
\mathfrak{B}_{\Lambda}$:
\begin{equation}\label{state<->measure}
\varphi^{(\Lambda)} (p)= \sum_{|\omega\rangle \langle \omega|
\leqslant p} \mu_{\Lambda} (\omega)
\end{equation}
Combining~\eqref{state<->measure} and
\eqref{consistency-operator-form-equation} and setting $a:=
|\omega_1\ldots\omega_m\rangle \langle \omega_m\ldots\omega_1|$ in
the latter, we obtain, for any $m, i \in \mathbb{N}$ and any
$\omega_{1}\ldots\omega_{m} \in {\mathcal{Z}}^{\Lambda (m)}$,
\begin{equation}\label{classical-consistency-cond}
\mu_{\Lambda (m)} (\omega_{1}\ldots\omega_{m})=
\sum_{\omega_{m+1}\ldots\omega_{m+i}} \mu_{\Lambda (m+i)}
(\omega_{1}\ldots\omega_{m} \omega_{m+1}\ldots\omega_{m+i})
\end{equation}
The equality~\eqref{classical-consistency-cond} is called the
(classical) {\em consistency} condition. Thus,
$\{\mu_{\Lambda}\}_{\Lambda \subset \zz}$ is a consistent family
of probability measures, and $\mu_{\Lambda}$ extends to a
probability measure  on $\bigl({\mathcal{Z}}^{\infty},
{\mathscr{P}}^{\infty}( {\mathcal{Z}}) \bigr)$ by the Kolmogorov
extension theorem\cite{kolmogorov}. The extended measure is
denoted by~$\mu$.
In fact, the tuple $(\mathfrak{B}_{\infty}, \varphi)$ and the
triple $\bigl({\mathcal{Z}}^{\infty},
{\mathscr{P}}^{\infty}({\mathcal{Z}}) \bigr)$ are just two
equivalent descriptions\cite{ruelle} of a given classical
stochastic process. This particularly implies the following
proposition.
\begin{propos}\label{abelian->classic}
If a state $\varphi$ on $\mathfrak{B}_{\infty}$ is stationary and
ergodic (weakly mixing or strongly mixing, respectively), then so
is the corresponding measure $\mu$ on
$\bigl({\mathcal{Z}}^{\infty},
{\mathscr{P}}^{\infty}({\mathcal{Z}}) \bigr)$. The converse is
also true.
\end{propos}
%\textbf{Proof of Proposition \ref{abelian->classic}:}\\
\begin{proof}
The result follows immediately from
Lemma~\ref{stationarity-operator-form},
Lemma~\ref{finite-ergodicity}, and  the
equality~\eqref{state<->measure}.
\end{proof}

%%%%%%%%%%%%%%%%%%%%%%%%%%%%%%%%%%%%%%%%%%%%%%%%%%%%%%%%%%%%%


\begin{thebibliography}{99}

\bibitem{barnum}
H.~Barnum, E.~Knill, and M.~Nielsen, ``On quantum fidelities and
channel capacities'', {\em IEEE Trans. Inform. Theory}, Vol.~46,
No.~4, pp.~1317--1329, July, 2000, {\em LANL e-print}
\url{http://xxx.lanl.gov/quant-ph/9809010}

\bibitem{berger}
T.~Berger, {\em Rate distortion theory; a mathematical basis for
data compression}, Englewood Cliffs, N.J., Prentice-Hall, 1971.

\bibitem{Q-McMillan}
I.~Bjelakovi\'c, T.~Kr\"uger, R.~Siegmund-Schultze, and A.~Szko\l
a, ``The Shannon-McMillan Theorem for Ergodic Quantum Lattice
Systems'', {\em LANL e-print}
\url{http://xxx.lanl.gov/math.DS/0207121}

\bibitem{bratteli-1}
O.~Bratteli, D.~Robinson, {\em Operator Algebras and Quantum
Statistical Mechanics I}, Springer-Verlag, New York, 1979.

\bibitem{bratteli-2}
O.~Bratteli, D.~Robinson, {\em Operator Algebras and Quantum
Statistical Mechanics II}, Springer-Verlag, New York, 1981.


\bibitem{petz}
F.~Hiai, D.~Petz, ``The Proper Formula for Relative Entropy and
its Asymptotics in Quantum Probability,'' {\em Commun. Math.
Phys}, Vol.~143, pp.~99--114, 1991.


\bibitem{king} C.~King, A.~Le\'sniewski, ``Quantum Sources and a
Quantum Coding Theorem'', {\em J. Math. Phys}, Vol.~39 (1),
pp.~88--101, 1998, {\em LANL e-print}
\url{http://lanl.arxiv.org/quant-ph/9511019}

\bibitem{kolmogorov} A.~N.~Kolmogorv, {\em Foundations of the Theory of
Probability}, Chelsea, Ney York, 1950.

\bibitem{kraus}
K.~Kraus, {\em States, Effects, and Operations}, Berlin, Germany:
Springer-Verlag,~1983.

\bibitem{ohya}
M.~Ohya, D.~Petz, {\em Quantum Entropy and its Use}, Springer,
Berlin, 1993.


\bibitem{mosonyi}
D.~Petz, M.~Mosonyi, ``Stationary Quantum Source Coding'', {\em
J.~Math. Phys}, Vol.~42, pp.~4857--4864, 2001, {\em LANL e-print}
\url{http://xxx.lanl.gov/quant-ph/9912103}


\bibitem{rudin1}
W.~Rudin, {\em Real and Complex Analysis}, McGraw-Hill, New York,
1987

\bibitem{rudin2}
W.~Rudin, {\em Functional Analysis}, McGraw-Hill, New York, 1973

\bibitem{ruelle}
D.~Ruelle, {\em Statistical Mechanics}, W.A. Benjamin, New York,
1969.

\end{thebibliography}
\end{document}